\documentclass[10pt]{article}

\usepackage{amsmath}
\usepackage{array}
\usepackage{appendix}
\usepackage{graphicx}
\usepackage{amsfonts}
\usepackage{amssymb}
\usepackage{mathrsfs}
\usepackage{yfonts}
\usepackage{euscript}
\usepackage{upgreek}
\usepackage{slantsc}
\usepackage{calligra}
\usepackage[T1]{fontenc}
\usepackage{epsf}
\usepackage{latexsym}

\usepackage{tipa}

\def\be{\begin{equation}}
\def\ee{\end{equation}}
\def\beq{\begin{equation}}
\def\eeq{\end{equation}}
\def\bea{\begin{eqnarray}}
\def\eea{\end{eqnarray}}

\def\foo{\footnote}

\def\!{\hspace{-1.6667em}}

\def\mC{\mbox{C}}   
\def\mD{\mbox{D}}

\def\mN{\mbox{N}}

\def\mS{\mbox{S}}

\def\md{\mbox{d}}

\def\mo{\mbox{o}}

\def\urho{{\underline{\rho}}}

\def\brho{\mbox{\boldmath$\rho$}}          
\def\fG{\mbox{\sffamily G}}

\def\fQ{\mbox{\sffamily Q}}
\def\fR{\mbox{\sffamily R}}
\def\fS{\mbox{\sffamily S}}

\def\fW{\mbox{\sffamily W}}

\def\bh{\underline{\underline{\mbox{h}}}  }            

\def\bp{\mbox{\bf p}}

\def\bQ{\mbox{\bf Q}}

\def\bh{\mbox{{\bf h}}}

\def\bq{\mbox{{\bf q}}}

\def\bfM{\mbox{{\bf \sffamily M}}}

\def\sa{\mbox{\scriptsize a}}

\def\se{\mbox{\scriptsize e}}

\def\si{\mbox{\scriptsize i}}

\def\sll{\mbox{\scriptsize l}}  
\def\sm{\mbox{\scriptsize m}}
\def\sn{\mbox{\scriptsize n}} 
\def\so{\mbox{\scriptsize o}}

\def\sr{\mbox{\scriptsize r}}
\def\st{\mbox{\scriptsize t}}

\def\sB{\mbox{\scriptsize B}}

\def\sG{\mbox{\scriptsize G}}
\def\sH{\mbox{\scriptsize H}}

\def\sJ{\mbox{\scriptsize J}}
\def\sK{\mbox{\scriptsize K}}

\def\sN{\mbox{\scriptsize N}}

\def\sR{\mbox{\scriptsize R}}
\def\sS{\mbox{\scriptsize S}}

\def\sW{\mbox{\scriptsize W}}

\def\sfA{\mbox{\sffamily{\scriptsize A}}}

\def\sfD{\mbox{\sffamily{\scriptsize D}}}

\def\sfF{\mbox{\sffamily{\scriptsize F}}}
\def\sfG{\mbox{\sffamily{\scriptsize G}}}

\def\sfK{\mbox{\sffamily{\scriptsize K}}}

\def\sfM{\mbox{\sffamily{\scriptsize M}}}

\def\sfW{\mbox{\sffamily{\scriptsize W}}}

\def\sfZ{\mbox{\sffamily{\scriptsize Z}}}

\def\sbfM{\mbox{\bf \scriptsize\sffamily M}}

\def\ta{\mbox{\tiny a}}

\def\te{\mbox{\tiny e}}

\def\ti{\mbox{\tiny i}}

\def\tl{\mbox{\tiny l}}

\def\tn{\mbox{\tiny n}}
\def\to{\mbox{\tiny o}}

\def\tr{\mbox{\tiny r}}

\def\ttt{\mbox{\tiny t}}   

\def\tG{\mbox{\tiny G}}

\def\tR{\mbox{\tiny R}}

\def\tfW{\mbox{\sffamily{\tiny W}}}

\def\lft{\mbox{\Large \sffamily t}}
\def\lt{\mbox{\Large $t$}}

\def\K{Kucha\v{r} }

\def\pa{\partial}
\def\d{\textrm{d}}

\def\5Star{\mbox{\Large$\star$}}              
\def\cr{\mbox{\scriptsize{\bf $\mbox{ } \times \mbox{ }$}}}

\def\sumi3{\sum\mbox{}_{\mbox{}_{\mbox{\scriptsize $i$=1}}}^3}

\def\sumIN{\sum\mbox{}_{\mbox{}_{\mbox{\scriptsize $I$=1}}}^{N}}
\def\sumj3{\sum\mbox{}_{\mbox{}_{\mbox{\scriptsize $j$=1}}}^3}
\def\sumk3{\sum\mbox{}_{\mbox{}_{\mbox{\scriptsize $k$=1}}}^3}

\begin{document}

\begin{titlepage}

\begin{center}

\Huge{\bf BACKGROUND INDEPENDENCE}

\vspace{.1in}

\normalsize

\vspace{.1in}

{\large \bf Edward Anderson}\foo{ea212@cam.ac.uk}

\vspace{.1in}

{\large {\em DAMTP Cambridge \normalsize}}.

\end{center}

\begin{abstract}

This paper concerns what Background Independence itself is (as opposed to some particular physical theory that is background independent).   
The notions presented mostly arose from a layer-by-layer analysis of the facets of the Problem of Time in Quantum Gravity. 
Part of this coincides with two relational postulates which are thus identified as classical precursors of two of the facets of the Problem of Time.  
These are furthemore tied to the forms of each of the GR Hamiltonian and momentum constraints.
Other aspects of Background Independence include the algebraic closure of these constraints, expressing physics in terms of beables, foliation independence as implemented by refoliation 
invariance, the reconstruction of spacetime from space.
The final picture is that Background Independence --- a philosophically desirable and physically implementable feature for a theory to have ---  
has the facets of the Problem of Time among its consequences. 
Thus these arise naturally and are problems to be resolved, as opposed to avoided `by making one's physics background-dependent in order not to have these problems'.
This serves as a selection criterion that limits the use of a number of model arenas and physical theories.  

\end{abstract}

\end{titlepage}

\section{Introduction}\label{Intro}

Background Independence is often (but not always) considered in attempts to combine General Relativity (GR) and Quantum Theory.  
Standard Quantum Theory is {\sl not} background-independent because it clings to either Newton's absolute structures (absolute time in particular) 
                                                                                 or to  the distinct absolute structures \cite{Kiefer} that Special Relativity replaced these with.  
I.e. Minkowski spacetime as a fixed background with its other type of privileged inertial frames and its timelike Killing vector.  
On the other hand, GR is not only a relativistic theory of gravitation but also Einstein's attempt to free Physics from background structures.
Here frames are only local and are determined by matter content, time is but a coordinate choice, and the generic GR spacetime has no Killing vectors of any kind.

Einstein's attempt did not directly address \cite{Ein} further relational Machian issues concerning time and space. 
However, it has since been determined that the theory of GR he arrived at in fact already manages \cite{B94I, RWR, Phan, ARel, FileR, AM13} to incorporate these. 
To proceed with explaining this, firstly the more standard and traditional dynamical/canonical formulation of GR -- Wheeler's Geometrodynamics \cite{ADM, Battelle} -- is required.
Here GR is reformulated as a dynamics of evolving spatial 3-geometries which stack together to form the standard formulation's spacetime.
This is usually described by evolving spatial 3-metrics which, unlike the description in terms of 3-geometries, have 3-coordinate (or, better, 3-diffeomorphism) redundancy 
but also the benefit of explicit computibility.  
Secondly, if one starts out from Leibniz--Mach relational first principles about time and space, one arrives at an equivalent variant of the Geometrodynamics formulation of GR.
(A similar analysis also applies \cite{FileR} if one considers instead Ashtekar variables-type dynamical formulations of GR \cite{AshtekarBook, RovelliBook, ThiemannBook} 
that have dominated the canonical, but not dynamical, GR literature since the late 1980's.)

Background Independence notions are often held to carry substantial physical, philosophical and conceptual content \cite{PC-Cont, Kuchar92I93, Intersect}. 
This is for reasons that go back to the centuries-old absolute versus relative (relational) motion debate \cite{AORM}.  
It should be said that the Background Independence attempted -- from Einstein to a) Ashtekar variables programs such as Loop Quantum Gravity \cite{RovelliBook, ThiemannBook} 
and b) the present paper -- is at the {\it metric level} \cite{MLBI, Kuchar92I93, Intersect}.
This is as opposed to topological-level Background Independence, for which humankind largely remains physically and mathematically unprepared to this day (see the Conclusion). 
Background Independence is also a stated goal in M-Theory \cite{M-Theory, SmoJe}, and furnishes one approach to Supergravity (particularly in canonical form) \cite{SUGRA-Can}.  
The above examples gives some idea of the opening sentence's `often'.
On the other hand, its `not always' is exemplified by covariant perturbative quantization of GR and of alternative theories (including perturbative String Theory), 
and by further theories mentioned in the Conclusion.  


Some precursors for the current paper's material and presentation are as follows.  
Most of it arose from analysing which facets of \K and Isham's Problem{\sl s} of Time \cite{Kuchar92I93, APoT, APoT2, FileR} in Quantum Gravity already occur at the classical level \cite{FileR}.  
This was found to recover \cite{ARel, FileR} many of the postulates, implementations, results and insights of Barbour-type Relationalism \cite{BB82, B94I, RWR, B11GrybTh, FileR}. 
This theory of Background Independence can also be seen as an extension --- from two postulates with implementations to {\sl nine} --- of a variant of Barbour's Relationalism: 
an extension that goes as far as covering all of the quantum aspects of the Problem of Time. 
I accredit Gryb \cite{Gryb10} for previously considered a theory of Background Independence based just on (a different variant of) Barbour's two relational postulates and 
implementations, with links to a small fraction of Problem of Time facets mentioned. 
However, the present work exceeds that by nine to two and by having comprehensive links to all of the known quantum and classically Problem of Time facets.  
Moreover, Barbour's Relationalism's two postulates proved to be highly unselective among the well-known alternative theories to GR \cite{Phan, Lan2}.
On the other hand, the Conclusion gives a number of examples of how the present paper's nine are substantially more selective in this manner.
Thus this paper provides a coherent enlargement of Barbour's work into a theory of Background Independence that 
1) is both usefully restrictive of widely proposed fundamental theories of Physics. 
2) It exposits the Problem of Time as a {\sl consequence} of the physically, philosophically and conceptually desirable option of directly implementing (metric) Background Independence. 


Note that 8/9ths of the possible alternative conceptual facets of the quantum Problem of Time already possess classical precursors \cite{FileR}.
Thus we start with the classical version.
The outline of the current paper is as follows.  
After the structural considerations and toy models outlined in   Sec 2, 
I cover Temporal Relationalism in                                Sec 3 
and Configurational Relationalism in                             Sec 4. 
These are the Barbour-type postulates and implementations. 
I then cover Constraint Closure in                               Sec 5, 
Expression in terms of Beables in                                Sec 6, 
GR-Type Spacetime Relationalism in                               Sec 7, 
Foliation Independence as attained by Refoliation Invariance in  Sec 8, 
Spacetime Reconstruction in                                      Sec 9.
These are the 7/9ths of the aspects of Background Independence that bring about 7/9ths of the facets of the Problem of Time [labelled exclusively by 1) to 9) in this article] 
that are to be faced by {\sl a local} resolution of the Problem of Time \cite{FileR, AHall, AMSS1-2}.
The remaining ninths are Globally Defined Specifications (which lead to the Global Problem of Time) and Uniqueness Specifications (which lead to the Multiple Choice Problem of Time). 
These are much less well understood in detail, as regards 1) the extent of the list of global sub-aspects of Background Independence (and of consequent Global Problem of Time). 
2) Also as regards addressing Uniqueness Specifications (and of the consequent Multiple Choice Problem) 
in a manner that is both mathematically sharp and applies to a wide range of models/theories.
So whilst I do give a conceptual outline of these in Secs 11 and 12, they largely remain among the frontiers of knowledge (see the Conclusion).  
Sec 10 concerns the classical Paths and Histories counterparts of Constraint Closure and Beables. 
Sec 12 also summarizes the connection between this paper's Background Independence postulates and Problem of Time facets.
These are labelled both with the names and conceptualizations used by Kucha\v{r} and Isham \cite{Kuchar92I93} and under my updated names for these \cite{APoT2, FileR}. 
Refer to Appendix A for a mosaic of figurelets that support the main text.  


\section{Structural preliminaries and toy models used}

\noindent I adopt the 
position that dynamics is primary (as opposed to e.g. spacetime being primary).  
I start with the notion of configuration $\bQ := Q^{\sfA}$ for $\sfA$ some indexing set.
I.e. the generalization of the `position' half of `positions and momenta' in the case of basic point-particle mechanics. 
Configuration space $\fQ$ is then the space of all possible values of the configurations that a system can possess. 
One is then to build composite objects out of the configurations.  
These include the {\it changes} of configuration $\md Q^{\sfA}$, actions, and the momenta $P_{\sfA}$ that correspond to the $Q^{\sfA}$.  
Some simple examples of configuration spaces are as follows; these also serve to introduce the toy models employed in the current paper.

\noindent Example 1) For the mechanics of $N$ particles in dimension $d$ in flat space 
the simplest (and redundant) configuration space is $\fQ(N, d) = \mathbb{R}^{Nd}$ \cite{Lanczos}.
This is the space of possible values of the particle positions $\bq^{\sfA} = \underline{q}^I$. 
[Underline denotes $d$-dimensional vector and $I$ a particle label running from 1 to $N$.] 

\noindent Example 2) Background-independent formulation of Mechanics -- {\bf relational particle mechanics (RPM)} \cite{BB82, QuadI, FileR} 
comes in scaled and pure-shape versions \cite{FileR}.
These already possess 5/9 of the aspects of Background Independence at the classical level and 6/8 at the quantum level \cite{FileR}, 
rendering them useful model arenas for quite a few Problem of Time investigations.  
The configuration space for this example is $\fR(N, d) = \mathbb{R}^{nd}$ for the scaled RPM of $N$ particles with the redundant translations with respect to an absolute origin trivially 
taken out by splitting out and discarding the centre of mass motion. 
This is the space of possible values of a basis of relative interparticle (cluster) separations, most conveniently Jacobi's, $\brho = \underline{\rho}^A$, 
for label $A$ running from 1 to $n := N - 1$.  
In 1-$d$ the scaled RPM has no other redundancies, so this manages to be a relational model with very simple and well-known configuration space mathematics.
(Thus it has well-understood dynamics and quantum theory. 
In this 1-$d$ case we drop the underline on the $\rho^A$.)

\noindent Each of the above two examples have an obvious Euclidean metric on configuration space (the $\mathbb{R}^{Nd}$ one rather than the spatial $\mathbb{R}^d$ one etc).

\noindent Example 3) For full GR, a redundant configuration space is $\mbox{Riem}(\Sigma)$ \cite{DeWitt67}.  
I.e. the space of all positive-definite 3-metrics $\bh = h_{ij}(x^k)$ on the fixed topological manifold $\Sigma$. 
The latter is taken here for simplicity to be compact without boundary, and which is interpreted as the theory's notion of space. 
[$x^k$ are the spatial coordinates and $i, j, k$ are spatial indices running from 1 to 3.]
Riem($ \Sigma$) furthermore possesses a physically-realized curved (-- + + + + +) signature per space point metric. 
I.e. the GR configuration space metric alias inverse DeWitt supermetric, ${\cal M}^{abcd} := \sqrt{h}\{h^{ac}h^{bd} - h^{ab}h^{cd}\}$.  

\noindent Example 4) A simpler subcase of Example 3) is {\bf minisuperspace} ($\Sigma$) \cite{Magic}: the space of homogeneous positive-definite 3-metrics on $\Sigma$. 
These are notions of space in which every point is the same.  
This possesses an overall (rather than per space point) curved (-- + + + + +) `minisupermetric'. 
Nested easier subcases within this are diagonal minisuperspace [flat 

\noindent (-- + +) minisupermetric]  and isotropic minisuperspace [flat (--) minisupermetric], 
for instance for $\Sigma = \mathbb{S}^3$ with standard hyperspherical metric.  
Minisuperspace \cite{Magic} has 7/9ths of the aspects of Background Independence classically and 8/9ths quantum mechanically, 
though quite a few of these imply {\sl very quickly resolved} Problem of Time facets by virtue of homogeneity.  
While the RPM and minisuperspace cases are simple to manipulate, they miss the subtleties specifically associated with diffeomorphisms \cite{Kuchar92I93}.  

\noindent Example 5) A first model in which these appear nontrivially is {\bf Halliwell--Hawking's model} of inhomogeneous perturbations about spatially-$\mathbb{S}^3$ minisuperspace.
This models a quantum-cosmological origin for the CMB hot-spots and galaxies.  
RPM's and minisuperspace complementarily support by one or the other having all {\sl other} Background Independence/Problem of Time aspects of this last model.   
Here one considers the first few (usually two) orders of the perturbation of the metric.
Each of these form a simplified configuration space in place of the full Riem($\Sigma$) that are currently under investigation.

\section{1) Temporal Relationalism}

This aspect of Background Independence adopts the Leibnizian `there is no time for the universe as a whole' principle \cite{B94I, FileR} at the primary level for closed universes.
At the classical level, this is mathematically implemented by the following. 

\noindent i) Postulate {\sl geometrical Jacobi--Synge type actions}.
These {\sl happen} to be parametrization-irrelevant (a mathematically straightforward but conceptually-curious dual property),
which is equivalent in turn to the earlier reparametrization-invariant conceptualization \cite{RWR} of this mathematical implementation.
My argument is this way round -- using and being named for the dual geometrical property -- on the grounds that it is a conceptual advance for background-independent physics  
to not name one's entities or techniques after physically-irrelevant properties.  
%

\noindent ii) Additionally, such actions are not to contain any extraneous time such as Newton's absolute time.
Nor are they to contain any extraneous time-like variables such as the ADM lapse of GR (named after the notion of `time elapsed').

\noindent The Euler--Lagrange formulation of Mechanics and the Arnowitt--Deser--Misner (ADM) formulation of GR immediately fail to obey ii). 
However, they are physically equivalent to the below formulations which do obey both i) and ii).\footnote{The equivalence of Jacobi actions to the more common Euler--Lagrange actions is 
by passage to the Routhian \cite{Lanczos, FileR}.
The equivalence of Misner's action to ADM's for minisuperspace is via Lagrange multiplier elimination (the minisuperspace subcase of the working in \cite{BSW}).}  

\noindent Example 1) Jacobi's action principle \cite{Lanczos} for spatially-absolute Mechanics.  

\noindent Example 2) Another case of Jacobi's action principle occurs for scaled 1-d RPM with translations trivially removed \cite{FileR}. 

\noindent Example 3) Misner's principle \cite{Magic} for minisuperspace GR.

\noindent All three of these actions principles are geometrical {\it geodesic principles} with respect to the physical line element $\d \widetilde{s}$.  
Alternatively, they are {\it parageodesic} (conformal to geodesic) principles with respect to the configuration space line element.  
These two alternatives can be summarized for all three examples at once as follows:    
\beq
S = \sqrt{2}\int \md \widetilde{s} := \sqrt{2}\int\d s\sqrt{W(\bQ)} \mbox{ } . 
\label{PI-geom-gestalt}
\eeq 
Here $\d s := ||\d \bQ||_{\sbfM}$ is the geometrical `kinetic arc element' built out of the configuration space metric $\bfM$. 
Also $W(\bQ)$ is the `potential factor'. 
This takes the form $E - V(\bq)$ for spatially-absolute mechanics, $E - V(\brho)$ for 1-$d$ scaled RPM or $R - 2\Lambda$ for minisuperspace (and, in Sec 4, generic, GR).
$E$ is total energy and $V$ is potential energy. 
$R$ -- generically of the form $ = R(x^k; h_{ij}]$ -- is the Ricci scalar corresponding to $h_{ij}(x^k)$ [though for minisuperspace it is just\footnote{I follow \K and Isham 
\cite{Kuchar92I93} in using  $( \mbox{ } )$ for function dependence, $[ \mbox{ } ]$ for functional dependence and $( \mbox{ } ; \mbox{ } ]$ 
for mixed function (before the ;) and functional (after the ;) dependence.
This then leaves me with $\{ \mbox{ } \}$ for ordinary brackets in equations.}
$R(h_{ij}$)] and $\Lambda$ is the cosmological constant.  
$\md \widetilde{s}$ itself is the physical arc element that is conformally related to the kinetic arc element by the overall conformal factor $\sqrt{W}$.  
Finally, `(para)geodesic' indeed implies `geometrical'.  
Moreover, in the given examples this refers to a particularly simple notion of geometry.
I.e. Riemannian geometry in the first two cases and semi-Riemannian -- indefinite-signature -- geometry in the last case.    
The geometrical arc-element for these all take the {\it quadratic} form of the square root of a sum of squares as given in the above expression for the action.
Synge's extension \cite{Lanczos, FileR} involves a wider range of physical theories being encoded by action principles built out of arc elements that encode more general notions of 
geometry than the above.

The following consequences ensue.  
I) Dirac \cite{Dirac} noted that primary constraints are implied by reparametrization-invariant actions.
This argument clearly carries over to geometrical actions that happen to be, dually, parametrization-irrelevant actions. 
Thus it can be recast as `primary constraints are implied by temporally-relational actions'.
Thus how the action for minisuperspace GR manages to encode the Hamiltonian constraint,  
\beq
{\cal H} := N_{ijkl}\pi^{ij}\pi^{kl}/\sqrt{h} - \sqrt{h}R = 0 \mbox{ } .  
\label{Ham}
\eeq 
This is relevant because in the more commonly used ADM approach, (\ref{Ham}) arises from variation with respect to the lapse, which in turn is necessarily absent from Misner's action 
for this to implement Temporal Relationalism.  
So there is an initial worry that the crucial constraint ${\cal H}$ may, somehow, have gone missing. 
However, Dirac's insight pin-points how upon formulating GR free of lapse, a distinct primary constraint route takes the place of the variation with respect to the lapse route to 
${\cal H}$ of the ADM formulation.
Thus in the background-independent paradigm for physics, the important Hamiltonian constraint of GR ${\cal H}$ arises directly from the demand of Temporal Relationalism. 
Indeed, furthermore, the precise form of the GR ${\cal H}$ is dictated by the precise (quadratic) way the action is built to be temporally relational.
This in turn induces ${\cal H}$'s quadratic dependence on the momenta, which, moreover, is well-known to cause the Schr\"{o}dinger picture manifestation of the 
most well-known quantum Problem of Time facet: the Frozen Formalism Problem.  
Thus indeed Temporal Relationalism the Background Independence aspect is a deeper and already classically-present replacement for the `Frozen Formalism Problem' Problem of Time facet.     
For spatially-absolute mechanics following from the Jacobi action, the quadratic energy constraint 
\beq
{\cal E} := \sumIN \bp_I^2/2m_I + V(\bq^I) = E
\eeq
plays an analogous role to ${\cal H}$, and the $I \rightarrow A$, $\bq \rightarrow \brho$ of this does likewise for 1-$d$ scaled RPM.  
In each case, the quadratic form of the geometrical arc-element in the action gives rise to a purely quadratic constraint `Quad' that then results in a quantum Frozen Formalism Problem. 
This is the first of a number of justifications for considering such Mechanics as useful model arenas for conceptually understanding some parts of the more elaborate case of GR itself.

II) The above primary-level timelessness from adopting the Background Independence Leibniz's Principle of Temporal Relationalism is, moreover, both already classically present 
and easier to reconcile with time nonetheless appearing to play a prominent role in the world around us.  
This reconciliation involves classical time being emergent at a secondary level via {\it Mach's Time Principle}: `time is to be abstracted from change'. 
Three distinct proposals for this then involve `any change' (Rovelli                                           \cite{Rfqxi}), 
                                               `all change' (Barbour                                           \cite{Bfqxi}) and my 
										   	  {\it sufficient totality of locally significant change (STLRC)} \cite{ARel2}.  
As detailed in \cite{ARel2}, all three of these proposals have some sense in which they are `democratic'   
However, only the last two take into consideration that `some clocks are better than others' is an essential part of accurate timekeeping \cite{Bfqxi}.
Additionally, only the first and the third are operationally realizable.\footnote{For STLRC, democracy is in all change having {\sl the opportunity} to contribute. 
Then only those changes whose contributions lead to effects above the desired accuracy are actually kept in practise, by which it also manages to be both operationally well-defined 
and a provider of accurate timekeeping.
Contrast this with `all change', for which, since some of the universe's contents are but highly inaccurately known or completely unknown, 
one can not include `all change' in accurate/practical calculations.}
%
Thus overall, STLRC wins out.  
The time abstracted from this is a {\sl generalization} of the astronomers' {\it ephemeris time} \cite{Clemence} that emphasizes that such a procedure is in practise {\sl local}. 
Thus I term it a `GLET', and posit the specialization of the Machian emergent time resolution to `GLET is to be abstracted from STLRC'.

For the actions in question, {\sl emergent Jacobi time} is such an emergent time.
This time arises as a simplifier \cite{B94I, FileR} of the change-momentum relations and Jacobi equation of motion.
These are the Temporal Relationalism complying equivalents\footnote{See \cite{AM13} for a summary of all such `TRiPoD' components 
(Temporal Relationalism incorporating Principles of Dynamics), or \cite{FileR, ABook} for an account of whichever of these.}
of velocity-momentum relations and Euler--Lagrange equations.  
It takes the form\footnote{I use `J' to denote `Jacobi', 
and the oversized notation $\mbox{\large $t$}^{\se\sm(\sJ)} = t^{\se\sm(\sJ)} - t^{\se\sm(\sJ)}(0)$ to incorporate selection of `calendar year zero', $t^{\se\sm(\sJ)}(0)$.}
\beq
\lt^{\se\sm(\sJ)}             =            \int \d s \left/ \sqrt{2 W(\bQ)} \right. \mbox{ } .
\eeq
At first sight, this is an `all change' resolution, but a careful look reveals that in practise it is indeed a `STLRC' resolution.
It amounts to a relational recovery of Newtonian, proper and cosmic times in the contexts in which one expects to have each such notion of time.  
Finally, suppose that one is in the presence of a heavy--light ($h$--$l$) split, such as for Cosmology for $h$ scale and $l$ small inhomogeneities.  
Then this scheme is only fully Machian once one passes from zeroth-order emergent times of the form 
\beq
\lt^{\se\sm(\sJ)}_0 = F[h, \d h] 
\eeq
to at least-first order emergent times of the form  \cite{ACos2}
\beq
\lt^{\se\sm(\sJ)}_1 = F[h, l, \d h, \d l] \mbox{ } .
\label{temJ1}
\eeq 
Such schemes indeed provide the classical $l$-degrees of freedom with the opportunity to contribute to the timestandard.

\section{2) Configurational Relationalism}\label{CR}

\noindent i) Now as well as starting with a geometrically-simple but redundant configuration space $\fQ$, one considers also a group $\fG$ of continuous transformations 
acting on the $\fQ$ that are taken to be physically irrelevant.
This encompasses both 

\noindent a) {\it Spatial  Relationalism}: translations and rotations relative to absolute space in Mechanics, or 3-diffeomorphisms in GR.
\noindent b) {\it Internal Relationalism}: a reformulation of the more familiar type of gauge theories from Particle Physics.
The meaning of Configurational Relationalism is that the true physics resides in the quotient space $\fQ/\fG$.
For example, for scaled RPM's this is {\it relational space} ${\cal R}(N, d) = \fR(N, d)/\mbox{Rot}(d)$. 
For pure-shape (i.e. additionally scale-free) RPM's, it is {\it shape space} $\fS(N, d) = \fR(N, d)/Dil \mbox{\textcircled{S}} \mbox{Rot}(d)$.  
Here Rot($d$) are rotations, Dil are dilations and \textcircled{S} denotes semidirect product. 
On the other hand, for GR $\fQ/\fG$ is usually taken to be Wheeler's \cite{Battelle, DeWitt67} 
                                       {\it superspace},           Superspace($\Sigma$) = Riem($\Sigma$)/Diff($\Sigma$).
Moreover cases can be made to consider {\it conformal superspace}, CS($\Sigma$)         = Riem($\Sigma$)/Conf($\Sigma$)    \mbox{\textcircled{S}}  Diff($\Sigma$) or 
                                       {\it CS + V},               \{CS + V\}($\Sigma$) = Riem($\Sigma$)/VPConf($\Sigma$)  \mbox{\textcircled{S}}  Diff($\Sigma$).  
Here, Diff($\Sigma$) are spatial 3-diffeomorphisms and Conf($\Sigma$) are conformal transformations.
V stands for `volume', i.e. the spatial volume of $\Sigma$ -- a merely global-valued variable.
Then VPConf($\Sigma$) are volume-preserving conformal transformations.
Conformal superspace is the space of conformal 3-geometries, so this case is a theory of {\it conformogeometrodynamics}.
\{CS + V\}($\Sigma$) adjoins to this the dynamics of a single volume degree of freedom.
Both of these spaces have close ties to prevalent approaches to the initial-value problem for GR \cite{ABFKO}.

Configurational Relationalism can be mathematically implemented in an indirect manner in a very wide range of circumstances by the following `$\fG$-act $\fG$-all' method. 
Given an object $O$ that corresponds to the theory with configuration space $\fQ$, one first applies a group-action of $\fG$ to this --- denoted $\stackrel{\rightarrow}{\fG}_gO$. 
Then one applies some operation $S_g$ that makes use of all of the $g^{\sfZ} \in \fG$ so as to cancel out the appearance of $g^{\sfZ}$ in the group action, 
such as summing, integrating, averaging or extremizing over $\fG$.

One example of such an implementation is for $O$ a classical action built upon $\fQ$ with application of the basic infinitesimal group-action to obtain
\beq
S_{\mbox{\scriptsize relational}} = \int \int_{\sN\so\sS} \d (\mN\mo\mS) \d_g s \sqrt{W(\bQ)}  
\mbox{ } , \mbox{ } \mbox{ } 
\d_gs := ||\d \bQ - \stackrel{\rightarrow}{\fG}_{\d g} \bQ||_{\sbfM}  \mbox{ } .  
\eeq
Then one extremizes over $\fG$ as per the variational principle now also including variation with respect to $g^{\sfZ}$.\footnote{`NoS' denotes each configurational entity's notion of space: 
3-space for field theories, whilst, for finite theories, we take $\int_{\sN\so\sS} \d (\mN\mo\mS) := 1$.}
%
This particular example is Barbour's {\sl Best Matching} \cite{ARel, B11GrybTh}. 
The extremization produces an equation, Lin$_{\sfZ}$, that, in the $Q^{\sfA}, \d Q^{\sfA}$ variables formulation, is to be solved for the auxiliary variables $\d g^{\sfZ}$ themselves 
and then substituted back into the action. 
This produces a final $\fG$-independent expression that {\sl directly} implements Configurational Relationalism.  
Note the extra $\int_{\Sigma}\d^3x\sqrt{h}$ factor in the GR case.

In more detail, this method initially looks to be `taking a step in the wrong direction' by extending $\fQ$ to the bundle $P(\fQ,\fG)$ that appends the further unphysical $\fG$-auxiliary 
degrees of freedom to the redundancies already present in $\fQ$.  
However, Lin$_{\sfZ}$, thus named for its linearity in the momenta, is, unlike the preceding quadratic constraints, uncontroversially a gauge constraint. 
Gauge constraints these use up {\sl two} degrees of freedom per $g^{\sfZ}$ degree of freedom. 
Thus each degree of freedom appended wipes out not only itself but also one of $\fQ$'s redundancies, so that one indeed ends up on $\fQ/\fG$ as required to implement Configurational 
Relationalism.

Note that the expression given involves formulating the actually-present auxiliary variables as $\d g^{\sfZ}$ rather than as, in the earlier literature \cite{BB82, RWR}, $g^{\sfZ}$.
This is necessary for these not to spoil the parametrization-irrelevance that implements Temporal Relationalism.
Moreover, then one might naively expect cyclic equations (Jacobi $\pa \, \widetilde{\d s}/ \, \pa \, \d g^{\sfZ} = \mbox{const}$ 
parallel of the conventional Lagrangian                          $\pa L/\pa \dot{g}^{\sfZ} = \mbox{const}$)           
in place of multiplier equations                         (Jacobi $\pa \, \widetilde{\d s}/ \, \pa g^{\sfZ}    = 0$ 
parallel of the conventional Lagrangian                          $\pa L/\pa \dot{g}^{\sfZ} = 0$).  
However, $g^{\sfZ}$ is entirely physically meaningless. 
Thus in particular its value at the end points (or, more generally, end NoS) of the variation are meaningless.
So it is to be subjected to free end NoS variation, and this fixes $C = 0$ at the end NoS, but $C$ is $NoS$-constant, so $C = 0$ everywhere. 
Thus using cyclic differentials $\d g^{\sfZ}$ in place of a Lagrange multiplier implementation of the $\fG$-auxiliaries leaves the equations of one's theory unchanged.

Additionally, in the presence of Configurational Relationalism, the preceding Sec's resolution of Temporal Relationalism picks up a complication:
\beq
\lft^{\se\sm(\sJ\sB\sB)} = \stackrel{\mbox{\scriptsize extremum $\d g \mbox{ } \in \mbox{ }$} \sfG}
                                                               {\mbox{\scriptsize of $S_{\tr\te\tl\ta\ttt\ti\to\tn\ta\tl}$}}                                                              
\left(                                                              
                                                               \int\d_{g} s \left/\sqrt{2\fW(\bQ)}\right. 
\right) \mbox{ } .  
\label{t-JBB} 
\eeq
JBB stands for `Jacobi--Barbour--Bertotti' \cite{BB82}, and the complication in question can be summed up as `more implicit via containing an extremum' and  
`formulated in terms of unphysical auxiliaries $\d g^{\sfZ}$'.
However, if one succeeds in carrying out Best Matching, the $\d g^{\sfZ}$ are replaced by an extremal expression solely in terms of $Q^{\sfA}, \d Q^{\sfA}$.  
By this, both aspects of this complication are washed away and one has an expression for $\lft^{\se\sm(\sJ\sB\sB)}$ just like the previous Sec's 
(but in terms of the {\sl reduced} configuration space's geometry).

\mbox{ } 

\noindent Example 1) The scaled relational mechanics case of (\ref{t-JBB}) follows from the relational action [see Fig \ref{RPM-coordi}.b)], 
\beq
S_{\sr\se\sll\sa\st\si\so\sn\sa\sll} = \sqrt{2}\int\d_Bs\sqrt{E - V} 
\mbox{ } , \mbox{ } \mbox{ } \mbox{ and }  \mbox{ } 
\d_B s := || \d\brho - \d B \cr \d \brho ||  \mbox{ } ,
\eeq
\beq
\lt^{\se\sm(\sJ\sB\sB)} = \stackrel{\mbox{\scriptsize extremum d$B$ $\mbox{ } \in \mbox{ }$ Rot(2)}}
                                   {\mbox{\scriptsize of $S_{\tr\te\tl\ta\ttt\ti\to\tn\ta\tl}$}}                                                              
\left(                                                              
                                                               \int \d_{B} s \left/ \sqrt{W(\bQ)}\right. 
\right) \mbox{ } .  
\eeq
The GR case of (\ref{t-JBB}) involves
\beq
S^{\sG\sR}_{\sr\se\sll\sa\st\si\so\sn\sa\sll} = \int \int_{\Sigma} \d^3x  \d_F s \sqrt{\sqrt{h}\{\mbox{Ric}(x; h] - 2\Lambda\}} 
\mbox{ } \mbox{ } \mbox{ and }  \mbox{ }  
\d_F s := ||\d_{F} h||_{\mbox{\boldmath\scriptsize $\cal M$}}
\mbox{ } ,  
\eeq
\beq
\lt^{\se\sm(\sJ\sB\sB)} = \stackrel{\mbox{\scriptsize extremum $\d F$ $\in$ Diff(3)}}{\mbox{\scriptsize  of $S^{\tG\tR}_{\tr\te\tl\ta\ttt\ti\to\tn\ta\tl}$}} 
\left(
\int    \d_F s
\left/
\sqrt{\mbox{Ric}(x; h] - 2\Lambda}
\right.
\right) \mbox{ } .
\eeq

{            \begin{figure}[ht]
\centering
\includegraphics[width=0.6\textwidth]{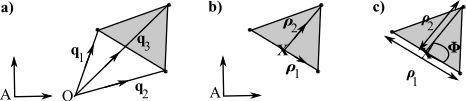}
\caption[Text der im Bilderverzeichnis auftaucht]{        \footnotesize{Progression of coordinate systems for the triangle. 
a) are particle position coordinates relative to an absolute origin and absolute axes.  
b) are relative Jacobi interparticle cluster separations; X denotes the centre of mass of particles 1 and 2; note that these coordinates are still relative to absolute axes.
Then the configuration space radius $\rho := \sqrt{\rho^2_1 + \rho^2_2}$.  
c) are scaled relational coordinates (i.e. no longer with respect to any absolute axes either).  
Pure-shape coordinates are then the relative angle $\Phi$ and some function of the ratio $\rho_2/\rho_1$.  
In particular, $\Theta := 2\,\mbox{arctan}(\rho_2/\rho_1)$ is then the azimuth to $\Phi$'s polar angle.} }
\label{RPM-coordi} \end{figure}          }
\noindent \mbox{ } \mbox{ } Configurational Relationalism the Background Independence aspect is related to one of the Problem of Time facets as follows.  
It is a threefold generalization of the Thin Sandwich Problem facet \cite{FileR}.  

\noindent I) Simply note that this is the subcase of this corresponding to $\fQ = \mbox{Riem}(\Sigma)$ and $\fG = \mbox{Diff}(\Sigma)$ {\sl is} the Thin Sandwich Problem when cast 
in terms of solving for an ADM or Baierlein--Sharp--Wheeler (BSW) \cite{BSW} multiplier coordinate shift $\beta^i$
See Fig 2.a) and b) for an outline of the Thin Sandwich prescription and its Thick Sandwich precursor; note that this Problem of Time facet is indeed usually considered at  
the classical level \cite{Kuchar92I93}.  
One can then pass to solving for the simultaneously Temporal Relationalism complying cyclic differential of a frame variable, $\d F^i$, 
without altering in any \cite{FileR} way the mathematical form of the thin sandwich equation \cite{BSW, TSP}.  

\noindent ii) Reconceptualize in terms of any $\fQ$ and a $\fG$ that is compatible with $\fQ$, and one has the Best Matching implementation of Configurational Relationalism

\noindent iii) Reconceptualize so that the elimination is at any level rather than at the Lagrangian ($Q^{\sfA}, \dot{Q}^{\sfA}$ variables) or Jacobian ($Q^{\sfA}, \d Q^{\sfA}$ 
variables) level, and one has the fully general `$\fG$-act, $\fG$-all' implementation of Configurational Relationalism.

Additionally, Best Matching is explicitly solvable for Example 1) 1- and 2-$d$ \cite{FileR, QuadI} (as is its pure-shape counterpart).  
By use of Kendall's Shape Theory \cite{Kendall}  the simplest configuration space geometries for 1- and 2-$d$ RPM's are $\mathbb{S}^{n - 1}$ and $\mathbb{CP}^{n - 1}$ (pure shapes).
Then by further use of the coning construction \cite{FileR}, $\mC(\mathbb{S}^{n - 1}) = \mathbb{R}^{n}$ and $\mC(\mathbb{CP}^{n - 1})$ (shapes and scale). 
The first three of these are very well known as geometries and as regards subsequent classical and quantum mechanics thereupon and supporting linear methods of Mathematical Physics.  
These render many QM and Problem of Time strategy calculations tractable and available for comparison with each other, which is a rarity in the latter field.  
Triangleland is further aided in this way by $\mathbb{CP}^{1} = \mathbb{S}^{2}$ and $C(\mathbb{CP}^1) = \mathbb{R}^3$ albeit the latter is not the usual flat geometry.  
It is, however, conformally flat \cite{FileR}.
The ensuing mathematics's simplicity, which extends even well into the usually complex rearrangements necessary for the investigation of Problem of Time strategies, 
is a major asset in this RPM model arena. 
This is because it secures many computational successes beyond the usual points at which these break down for full GR/many other toy models.  
Pure-shape RPM configuration spaces are analogous to conformal superspace (CS) \cite{CS} for GR, 
and scaled ones to Wheeler's superspace in one sense and to CS + Volume \cite{ABFKO} in another sense.
\cite{FileR, QuadI} demonstrated solvability, by mixture of basic maths and interdisciplinarity with statistical theory of shape, 
Molecular Physics and a few other areas for $N$-a-gonlands (Particle Physics, instantons).

For use below, scaled triangleland's reduced configuration spaces possesses Cartesian coordinates that are non-obviously related to the merely-relative Jacobi coordinates. 
I.e. the `Dragt coordinates' \cite{FileR} 
\beq
Dra_1 = 2\urho_1 \cdot \urho_2 \mbox{ } , \mbox{ } \mbox{ }
Dra_2 = 2\{\urho_1  \cr  \urho_2\}_3 \mbox{ } , \mbox{ } \mbox{ }
Dra_3 = {\rho_2}^2 - {\rho_1}^2 \mbox{ }  
\eeq
(the form of which arises by what is more widely known as the `Hopf map').  
These are all cleanly interpretable as the product of a scale factor $I$ and a lucid shape quantity.
Respectively, these shape quantities are an anisoscelesness, four times the mass-weighted area of the triangle and the ellipticity (difference of partial moments of inertia).

\noindent ii) Much as Temporal Relationalism came with a second statement about no extraneous times, 
Configurational Relationalism comes with a second statement of no extraneous spatial metric.  
[In all cases in this paper, accompanying `no extraneity' statements are metric geometry extraneities, the lapse itself being interpretable as playing the role of a 1-metric of time.]

This paper now branches, firstly into canonical bracket structure in Secs 5 and 6 and then into spacetime structure in Secs 7, 8 and 9.

\section{3) Constraint Closure}

\noindent As well as the $Q^{\sfA}$ and $P_{\sfA}$ introduced above, one requires to endow the combined space of these 
with the further Poisson brackets structure, $\mbox{\bf \{} \mbox{ } \mbox{\bf ,} \mbox{ } \mbox{\bf \}}$.

Then the constraints have Poisson brackets among themselves. 
For all of RPM's, GR and minisuperspace, the constraints already found close algebraically at the classical level without further issue.  
Cases with Best Matching explicitly attained are aided in the matter of closure by a) having less constraints to make brackets out of.
b) By single finite-theory classical c-number constraints always closing with themselves by symmetric entries into an antisymmetric bracket.  
In more detail, for GR, the constraints' Poisson brackets form the Dirac Algebroid\footnote{This is given here with the TRiPoD-consistent variant form 
for the smearing functions, d$L^i(z)$ and d$M^i$ for ${\cal M}_i$ and d$J$, d$K$ for ${\cal H}$. 
$({\cal C}_{\tfW}|\d A^{\tfW}) := \int d^3z \, {\cal C}_{\tfW}(z^i; h_{jk}] \, \d A^{\tfW}(z^i)$ is an `inner product' notation for the smearing of a $\fW$-tensor density valued 
constraint ${\cal C}_{\tfW}$ by an opposite-rank differential $\fW$-tensor smearing with no density weighting, $\d A^{\tfW}$.}
\be
\mbox{\bf \{} (\mathcal M_i | \d L^i \, ) \mbox{\bf ,} (\mathcal M_j | \d M^j \, ) \mbox{\bf \}} =  (\mathcal M_i | \, [\d L , \d M ]^i )  \, ,
\label{M,M}
\ee
\be
\mbox{\bf \{} (\mathcal H | \d K \, ) \mbox{\bf ,} (\mathcal M_i , \d L^i \, ) \mbox{\bf \}} = (\pounds_{\d L} \mathcal H | \d K \, ) \,     , 
\label{H,M}
\ee
\be 
\mbox{\bf \{} (\mathcal H | \d J \, ) \mbox{\bf ,} (\mathcal H | \d K \, )\mbox{\bf \}}  = ( \mathcal M_i |  {\d J} \, \overleftrightarrow{\pa}^i {\d K})  \, , 
\label{H,H}
\ee
where $X \overleftrightarrow{\pa}^i Y := ( \pa^i Y ) X - Y \pa^i X$ (a notation familiar from QFT).

The first bracket means that the 3-diffeomorphisms on a given spatial hypersurface close in the usual way as a Lie algebra.
[I.e. the right-hand side is of the form of a sum of constraints each multiplied by a {\it structure constant}.]    
\noindent The second bracket means that ${\cal H}$ is a good object (a scalar density) under 3-diffeomorphisms.
\noindent The third bracket, however, is more complicated in both form and meaning.
For now, I note that the presence of $h^{ij}(h_{kl})$ on its right-hand side is a structure function. 
Thus this bracket renders the overall algebraic structure of the brackets more complicated than a Lie algebra.
It is, rather, an algebroid; in particular, it is the {\it Dirac algebroid}.  
This is entirely unlike the {\sl unsplit} GR's 4-diffeomorphisms of spacetime $M$ which form a genuine Lie algebra just like the 3-diffeomorphisms do.
The Dirac algebroid is a highly diffeomorphism-specific structure. 
In the case of minisuperspace, it has the last right-hand side zero by homogeneity ($D_i$ annihilates everything).  
I also note that further interpretation of the third bracket is largely the basis of Secs 8 and 9.

Constraint closure is indeed a necessary check since the constraints that one readily finds can sometimes imply further conditions.
Dirac \cite{Dirac} gave a systematic treatment for this at the classical level, in terms of Poisson brackets $ \mbox{\bf \{} \mbox{ } , \mbox{ } \mbox{\bf \}}$ 
on the space of the $Q^{\sfA}$ and $P_{\sfA}$.   
One's theory's constraints' Poisson brackets are capable  of the following \cite{Dirac}.

\noindent i)  Producing further                          constraints.

\noindent ii)  Or of    diagnosing these or pre-existing constraints                          as in fact being second-class. 

\noindent iii) Or of    producing a further              type of equation called `specifiers' that fix the forms of some of the theory's auxiliary quantities.
Further constraints thus produced may mean that the $\fG$ one intended to enforce as physically meaningless is rejected or enlarged by the physics of one's theory.  
Second-class constraints thus diagnosed do not have the property of using up a total of two degrees of freedom per auxiliary $\fG$-degree of freedom.  
Thus the presence of such may be sending the physical theory to other than the intended quotient.

These issues are met by some of the following means. 

\noindent i) Sometimes one's attempted physical theory has to be abandoned. 
Not all theories one can write down are consistent!
Dirac illustrated this with the Lagrangian $L = x$, whose Euler--Lagrange equation reads 0 = 1...

\noindent ii) Sometimes the $\fG$ one initially selected has to be abandoned.  
For instance, attempting GR-like metrodynamics instead of geometrodynamics turns out to be impossible because ${\cal M}_i$ turns out to be an integrability of ${\cal H}$.  

\noindent iii) One can try to deal with any second-class constraints one finds by gauge-fixing them.
Or by extending the phase space. 
Or by replacing the Poisson bracket with a Dirac bracket that factors out the second-class constraints \cite{Dirac, HT}.

\section{4) Expression in terms of Beables}

\noindent Having found constraints and introduced the classical Poisson brackets structure, one is additionally free to ask about which objects have zero Poisson brackets with the 
constraints. 
These objects, termed {\it observables} or {\it beables}, are more useful than just any $Q^{\sfA}$'s and $P_{\sfA}$'s or functionals of these.
This usefulness follows from their containing physical information only.  
At the very least this is required to be present in final answers to questions posed to a physical theory.
My particular program emphasizes {\it beables}: quantities that just `are', rather than carrying any connotations of `observing'.
At the classical level, this concept is already more suitable from a cosmological perspective \cite{Bell}.  
One still has an ambiguity as to which notion of beables to use. 

\noindent {\it Dirac beables} are required to have zero brackets with all of the constraints: $\{{\cal C}_{\sfW}, D_{\sfD}\} = 0$.

\noindent {\it \K beables} are less stringent in being required to have zero brackets only with the linear constraints: $\{\mbox{Lin}_{\sfZ}, K_{\sfK}\} = 0$.

\mbox{ } 

\noindent Note 1) \K beables are the end-product of succeeding to find a full algebra of gauge-invariant quantities. 
The `Poisson brackets' in use might in fact belong to an extended phase space, or be Dirac brackets \cite{Dirac} constructed so as to remove second-class constraints.   
The gauge-invariant quantities in question correspond to the successful realization of Configurational Relationalism's $\fG$ as a means of passage to a less redundant quotient 
configuration space $\fQ/\fG$.  

\noindent Note 2) Immediately from the Jacobi identity on two constraints and one beable,  the `Poisson bracket' of two beables is itself a beable. 
Thus each of Dirac and \K beables constitute a further classical algebra. 

\mbox{ } 

\noindent An alternative choice suggested are  {\it Rovelli partial observables} \cite{RovelliBook}, which have no such requirements and are not physical quantities per se, 
though correlations between pairs of such are physical quantities.  
One often imagines each as measured by a localized observer, hence this concept is usually expressed as observables rather than as beables. 
Moreover one can conceive of beables with all the other Rovellian properties, that are e.g. locally decohered by processes bereft of the trappings of measurements and observers.
[Rovelli's scheme also makes use of `total observables' that are similar to Dirac's notion of observables.] 
\noindent See Secs \ref{SR}, \ref{PH} for yet more notions of beables/observables.  

\mbox{ } 

\noindent The connection with a Problem of Time facet is as follows. \K and especially Dirac beables are hard to find, and even more especially at the quantum level.   
This accounts for the Problem of Beables (more often called `Problem of Observables') facet being a problem. 
That it may be a forteriori a Problem of Time facet is tied to commutation with ${\cal H}$ also possibly carrying connotations of constants of the motion, i.e. the conundrum that 
one's desirable beables might also all be frozen in time.  
Additionally, at least directly formulated Dirac beables have to contend with possible nonexistence, even at the classical level, due to nonintegrability/chaos. 

\mbox{ } 

\noindent Some benevolent cases are as follows.  

\noindent i) Trivial Configurational Relationalism (or resolved Best Matching) readily imply possession of a full set of classical \K beables. 
E.g. explicitly for the relational triangle, 
\beq
\mbox{K} = {\cal F}[Dra, \Pi^{Dra} \mbox{ alone}] \mbox{ } , 
\eeq
for $Dra$ the lucid scaled-up shape quantities introduced in Sec 3 with conjugate variables $\Pi^{Dra}$.
Such shape quantities are also available for the general 1- and 2-$d$ RPM and for diagonal minisuperspace models.

\noindent ii) In the case of trivial Configurational Relationalism, Halliwell provided \cite{H03, H09H11} an indirect prescription for Dirac beables.
I.e. objects commuting with Quad also, which I further promoted to the case of resolved Best Matching too \cite{AHall} (constructed for the relational triangle).  

\noindent The Problem of Beables then consists of finding objects which brackets-commute with all the constraints (Dirac beables) or perhaps just with the linear constraints.

\noindent iii) I distinguish between specific and merely formal resolutions of the Problem of \K Beables. 
E.g. for the relational triangle, one has a specific set of shape quantities (Sec \ref{CR}). 
Whereas for GR one can only talk formally in terms of the spatial 3-geometries and their conjugates.

\section{5) Spacetime Relationalism}\label{SR}

i) Now as well as considering a spacetime manifold $M$, consider a $\fG_{\sS}$ of irrelevant transformations acting upon $M$.    

\noindent ii) Also do not consider any extraneous spacetime structures, in particular indefinite background spacetime metrics.

\noindent For GR, $G^{\prime}$ = Diff($M$), which, as mentioned above, {\sl do} form a Lie algebra, 
\be
|[ (\mD_{\mu}|X^{\mu}), (\mD_{\nu}|Y^{\nu}) ]| = (\mD_{\gamma}|\,\,[X, Y]^{\gamma}) \mbox{ }  . 
\ee
There is still an issue as to what role the Diff($M$) generators $\mD_{\mu}$ play here. 
Unlike with Sec 4's Diff($\Sigma$) auxiliaries, these $\mD_{\mu}$ are not conventionally associated with dynamical constraints.      
Nor is the above classical realization of a Lie bracket conventionally taken to be a Poisson bracket. 
Because of that, there is conventionally no complete spacetime analogue of the previous Sec's notion of beables/observables.
However, the notion of Diff($M$) invariant quantities given by objects $S_{\sfF}$ such that 
\beq
|[ (\mD_{\mu}|X^{\mu}), (S_{\sfF}|Z^{\sfF}) ]|  = 0
\eeq
does remain useful and physically meaningful.
The Jacobi identity applying to all Lie brackets, these $S_{\sfF}$ are themselves also guaranteed to close as a Lie algebra.

\noindent Moreover, the well-known Einstein--Hilbert action of the spacetime formulation of GR is built out of a good spacetime scalar. 
Thus there are no manifest corrections at the level of the action in this case. 
[The $\d g$'s that entered Sec 4's actions did so via their split-off kinetic terms not constituting good $\fG$ objects without such corrections.]
On the other hand, corrected-derivative entities analogous to Best Matching do still exist for spacetime, even though these are now to have an interpretation in terms of an 
auxiliary rather than physically-realized 5-manifold.
E.g. GR perturbation theory can be cast as an unphysical but technically-useful 5-$d$ stack of spacetime 4-geometries that are inter-related via Lie derivative terms  \cite{5-Lie}.
with respect to, now, a {\sl spacetime 4-vector}.  
Such has the status of a {\sl resolved} problem at the classical level.

More usually, one makes a choice to work with one of split or unsplit spacetime.
However, a few approaches to Background Independence and Quantum Gravity do {\sl combine} spacetime and split spacetime concepts.
Thus they manifest all of Temporal, Spatial and Spacetime Relationalism at once. 
See Sec \ref{PH} for examples.

\section{6) Foliation Independence and Refoliation Invariance}

\noindent {\it Foliation Dependence} is a type of privileged coordinate dependence. 
This runs against the basic principles of what GR contributes to Physics.
Thus {\bf Foliation Independence} is an aspect of Background Independence.

\noindent{\bf Refoliation Invariance} is that evolving via the dashed surface and via the dotted one in Fig 2.e) give the same physical answer.
\noindent This is also known as {\it path independence}.  
This refers to paths in GR's configuration space, each of which represents a distinct sequence of spatial 3-geometries.
Refoliation Invariance is a property manifested by the classical canonical formulation of GR that guarantees Foliation Independence in this context.  
This occurs via Fig 2.e) being guaranteed by the Poisson bracket (\ref{H,H}) closing it up modulo a mere diffeomorphism on the second hypersurface 
as per the third figurelet in Fig 2.g) (an insight of Teitelboim \cite{T73}).  
\noindent It is therefore a diffeomorphism-specific resolution.

By this property, GR spacetime is not just a strutting together of spaces.
Rather, it manages to be many such struttings at once in a physically mutually consistent manner.
Even the relational form of mechanics fails to manifest this property; this is one of the key limitations of the RPM model arena.
On the other hand, minisuperspace is often considered in terms of the privileged foliation by spatially homogeneous surfaces, which greatly simplifies this facet \cite{AMSS1-2}. 
Both these limitations are to be expected since Refoliation Invariance is a diffeomorphism-specific issue.
Moreover, just as Dirac pointed out that foliation of Minkowski spacetime by arbitrary spatial 3-surfaces does involve diffeomorphism-specific issues, 
a similar treatment does also exist for minisuperspace cosmologies.
%

Finally, Hojman-Kucha\v{r}-Teitelboim \cite{HKT76} obtained the form of ${\cal H}$ from less assumptions than forming the ADM split of spacetime GR. 
They proceeded via the deformation algebroid for a hypersurface taking the form of the already-known Dirac algebroid.  
This does however still presuppose (embeddability into) spacetime.

\section{7) Spacetime Reconstruction}

\noindent On the other hand, Barbour--Foster--\'{o} Murchadha began a program \cite{RWR} (the most recent form of which is in \cite{AM13}) 
deriving the form of ${\cal H}$ without presupposing spacetime structure.  
They assumed Temporal and Configurational Relationalism instead.
Thus their approach additionally amounts to a classical spacetime reconstruction (from space).

Another kind of Spacetime Reconstruction would be from a discrete ontology rather than a continuum [whether for spacetime or for space: Fig 2.f)].
We shall see in Sec \ref{QM-Sec} that Spacetime Reconstruction is further motivated at the quantum level, but it does already admit a classical precursor. 
In general, if classical spacetime is not assumed, one needs to get it back in a suitable limit. 
This can be hard; in particular, the less structure is assumed, the harder a venture it is.

In the case of assuming a space continuum, mere consistency imposed by the Dirac procedure whittles down a general ansatz to one of four alternatives.
I.e. Lorentzian, Galilean, Carrollian relativity and CMC (constant mean curvature) slicing \cite{RWR, AM13}.  
These arise together as the different ways to kill off an obstruction term that is a product of four factors.
It is novel for such an alternative to arise from Principles of Dynamics considerations. 
[This is as opposed to the historical dichotomy between universal local Galilean or Lorentzian relativity.] 
It is furthermore intriguing that it gives the CMC slicing option -- familiar from York's work \cite{CS, ABFKO} on the initial value problem -- 
as an option on the same kind of footing as this dichotomy.

\section{Path and Histories notions of Beables}\label{PH}

One might guess that $S_{\sfF}$ could be viewed as beables in the distinct sense of Bergmann. 
This is because both are gauge-invariant path notions.  
A path in geometrodynamical phase space corresponds to sweeping out a slicing of a spacetime, encoding both the sequence of 3-metrics and the extrinsic curvature of each of these.
However, Bergmann and Komar \cite{BK71} note that these have invariances much larger than Diff($M$), and their position is that the largest group of invariances is to be taken. 
Thus their $S^{\sB\sK}_{\sfM}$ are distinct from $S_{\sfF}$ by their connection to a larger group Digg($M$) -- the diffeomorphism induced gauge group -- in place of Diff($M$).

On the other hand, in classical Histories Theory \cite{Hist}, classical paths in configuration space (which represent particular slicings of spacetime)
%
%
are elevated to have a similar canonical status to ordinary configurations. 
I.e. they are taken to possess conjugate {\it histories momenta} and {\it histories brackets}: Poisson brackets on the ensuing histories phase space.
There is then a notion of histories constraints, ${\cal C}^t_{\sfW} = \langle \mbox{Quad}^t, \mbox{Lin}^t_{\sfZ} \rangle$ and of histories Dirac and Kucha\v{r} beables, 
\be
\mbox{\bf \{} \mbox{Lin}^t_{\sfZ}, K^{t}_{\sfK} \mbox{\bf \}}_{\sH} = 0 \mbox{ } , 
\ee
\be
\mbox{\bf \{} {\cal C}^t_{\sfW}, D^{t}_{\sfD} \mbox{\bf \}}_{\sH} = 0 \mbox{ } ,
\ee
each of which closes as an algebra by the corresponding Jacobi identity.  
Due to having this interpretation pinned upon them, histories beables are distinct from each of spacetime beables and Bergmann beables.

I finally comment on approaches that are concurrently covariant {\sl and} canonical. 
Such have been outlined by Savvidou \cite{Savvidou04ab}, treated for toy models by Kouletsis and Kucha\v{r} \cite{KK} and in greater generality (but only at the classical level) 
by Kouletsis \cite{Kouletsis08}.  
These would be expected to concurrently exhibit Temporal, Configurational and Spacetime Relationalism cast in terms of histories, 
and both Configurational and Histories Beables notions.
This issue is investigated further in \cite{POB}.

\section{8) Globally Defined Specifications (in outline)}

\noindent It would be preferable if all entities and strategies used are globally are well-defined, avoidable or accountable for in terms of a meaningful physical interpretation.  
A few examples of consequent classical-level Global Problem of Time subfacets impinging upon the current paper's exposition of Background Independence are as follows.
(I shall give a substantially more complete list of these in \cite{AGlob}, alongside various classification schemes. 
Local in space or in time itself? 
A simple issue of patching together maps, or a far less simple issue of patching together partial differential equations' solutions or unitary evolutions?)

\noindent i) If there is a time-function, is it defined everywhere in space and in time itself? 

\noindent ii) If there are true dynamical degrees of freedom, are these defined everywhere in space and in time?

\noindent iii) How does one deal with the emergent time and the best-matching procedure globally? 

\noindent iv) Do definitions of beables hold globally, and, if not, how does one move from one set to the other? 

\noindent v) The refoliation-invariance and spacetime reconstruction demonstrations remain only local.

\section{Quantum counterparts and their links to Quantum Problem of Time facets}\label{QM-Sec}

\noindent 1) {\bf Frozen Formalism Problem}.         
The quadratic Hamiltonian constraint of GR gives a time-independent Schr\"{o}dinger equation at the quantum level in a situation in which one might expect a time-dependent one.
This is well-known to be a consequence of ${\cal H}$ being quadratic and not linear in the momenta. 
What is hitherto less well-known is Sec 3's exposition that this in turn is a consequence of the form of relational action and thus of the {\bf Temporal Relationalism} 
that this implements.

\noindent Note 1) The Machian classical emergent time resolution breaks down at the quantum level: it does not unfreeze the GR quantum wave equation.  
However, this resolution can now be replaced by the equally-Machian semiclassical resolution, which additionally differs in the detail of its corrections for obvious STLRC reasons. 
I.e. now somewhat distinct quantum $l$-changes have an opportunity to contribute in place of classical $l$-changes.  
For instance, in the semiclassical approach \cite{HallHaw, KieferBook} wavefunctions of Born--Oppenheimer--WKB form 
\beq
\Psi = \mbox{exp}(iS(h)/\hbar)|\chi(l, h)\rangle \mbox{ } , 
\eeq
$\lt^{\se\sm(\sW\sK\sB)}$ can be interpreted as \cite{FileR, ACos2} a Machian emergent semiclassical time. 
To zeroth order, this coincides with $\lt^{\se\sm(\sJ)}_0$, but both fail to be Machian on account of allowing neither classical nor semiclassical $l$-change to contribute.   
To first order, however,
\beq
\lt^{\se\sm(\sW\sK\sB)}_1 = F[h, l, \d h, |\chi(l, h)\rangle] \mbox{ } ,
\eeq 
which is clearly distinct from (\ref{temJ1}).  

\noindent Note 2) The wave equation does not suffice to obtain physical answers, which require also an inner product input.  
This is tied to time issues via conservation of probability/unitarity and it is in general unclear how to select an inner product with suitable properties for Quantum Gravity.
I.e. the {\bf Inner Product Problem} subfacet of the Frozen Formalism Problem.  

\noindent Note 3) One's quantum formalism may include further objects needing to be cast into temporally-relational form, such as path integrals and quantum operators.
%

\mbox{ } 
															
\noindent 2) {\bf Quantum Configurational Relationalism}. 
If one succeeds in reducing out $\fG$ at the classical level, one may not need to face a $\fG$ again at the quantum level. 
On some occasions however [see 3) below], the quantum theory requires a distinct $\fG^{\prime}$ .
Were that to happen, or were one unable to reduce $\fG$ out at the classical level, one could continue to use the indirect $\fG$-act, $\fG$-all method for all subsequent 
objects required by one's theory.
For instance, for classical notions of distance, classical and quantum notions of information and quantum wave equations and operators \cite{FileR}.
In the semiclassical case, were this to happen, one needs to replace not only the integrand of (\ref{t-JBB}) with operator-ordering- and expectation-term corrections.
But also consider what to replace the classical action within the accompanying extremization condition. 
The first of these is considered in \cite{ACos2} and the second will be written up in v4 of \cite{FileR}.

\mbox{ } 
			
\noindent 3) {\bf Functional Evolution Problem}.    
This is \cite{Kuchar92I93} whether the Hamiltonian and momentum constraints are all that one needs at the quantum level by virtue of closing and of not producing anomalies.
However, this name for the problem refers only to the field-theoretic case.
Including the finite case also involves naming it after the portmanteau of the partial and the functional derivatives \cite{FileR}.
But even then it is still only a quantum-level name, whereas the term {\bf Constraint Closure Problem} extends it to include its classical counterpart too.

At the quantum level, this involves the following further issues.

\noindent A) The classical bracket is replaced by the quantum commutator.
However, this does not always obey the same algebraic structure as the classical bracket, for starters for global reasons \cite{I84}.  

\noindent B) Additionally, the classical constraints are replaced by operator-valued quantum constraints. 
However, this is subject to operator-ordering ambiguities, and to possible regularization and functional-equation well-definedness issues. 
Furthermore, how these are operator-ordered will almost always affect the form of the quantum commutator brackets, thus feeding into both A) and C).  

\noindent C) {\it Anomalies} can arise.
These are a type of brackets algebra obstruction that specifically alters the classical symmetry group at the quantum level. 
Thus it can be that a suggested use of $\fG$ as a group of irrelevant transformations is accepted by one's classical theory only to be rejected at the quantum level.  

\noindent Note 4) A), B) and C) need not always be subfacets of the Constraint Closure Problem because some instances of them are unrelated to time, space or frames.
Thus in this case these play no part in Background Independence or the Problem of Time.  

\mbox{ } 
			
\noindent 4) The {\bf Problem of Observables} \cite{Kuchar92I93} has been argued to be better named and conceived of as the {\bf Problem of Beables}.
This is now not only as regards the cosmological setting but also as regards the non-existence of ordinary QM's notion of observers 
or measuring apparatuses in many relevant universe regimes. 
The word `beable' was indeed always meant \cite{Bell} to carry further realist-interpretation connotations at the quantum level.
Variants of such interpretations that remain alive and well include 1) decoherence paradigms, 
2) many-worlds interpretations, 3) Histories Theory and 4) the contextual-realist approach of Doering and Isham \cite{ID}.  
These all have variants that can be used for whole-universe Quantum Cosmology.

The quantum Problem of Beables thus involves three changes as compared to its classical precursor. 
I.e. it now involves zero commutator brackets between operator-valued versions of both the constraints and the beables.
These further assume the end-product of Background Independence postulate 3) has been attained.
Plus one now has to worry about, firstly, how to select a subalgebra of one's classical beables to promote to quantum beable operators.
Secondly, about how one is to operator-order these so as to ensure the following.  

\noindent i)  The beables close among themselves.

\noindent ii) The beables succeed in forming zero brackets with the final selection of operator-ordered and regularized quantum constraints.  

\noindent The quantum Problem of Beables is consequently that \K and especially Dirac beables are in general even harder to find at the quantum level than they were at the classical level.  

\mbox{ } 

\noindent 5) {\bf Spacetime Relationalism}.           
This does not occur if one simply splits spacetime and works canonically. 
However, if an unsplit spacetime formulation is used, one has the advantages of no frozen wave equation or inner product problem or foliation issues to worry about. 
At the quantum level, such unsplit approaches furthermore make use of Feynman's path integral technique. 
Whilst very successful in QFT, for Quantum Gravity these face their own set of problems. 

\noindent I) In place of an Inner Product Problem, there is a {\bf Measure Problem}.
Moreover, this is diffeomorphism-specific and thus is itself directly a Problem of Time facet resulting from the Spacetime Relationalism Background Independence aspect.  

\noindent II) The gravitational path integral is, at the very least, problematic as regards being well-defined. 
(Discrete approaches can overcome this aspect).

\noindent III) The analogue of the Wick rotation familiar from QFT (`from imaginary time to real time') in general goes awry in curved spacetimes. 
This is due to ambiguities arising as regards which contours to use \cite{HL90}.

Also, QFT is free to rely on the canonical approach as regards computing what its Feynman rules are. 
Thus ordinary QM `in terms of path integrals' can sometimes in fact be a combined path-integral {\sl and} canonical approaches.
Then in Quantum Gravity, such a combined scheme would pick up the problems of {\sl both} approaches.

Finally, one might use quantum-level Histories Theory instead, i.e. with more structures than are usually assumed in a path integral approach.  
This changes some of the mathematics involved, and is at least one way of attempting combined spacetime-and-canonical schemes.  
However, such programs remain largely incomplete (\cite{Kouletsis08} is classical only).

\mbox{ }

\noindent 6) {\bf Foliation Dependence Problem}.  
There is no known general counterpart at the quantum level of the classical Refoliation Invariance guarantee of the third figurelet of Fig 2.g), not even at the semiclassical level.  
[The present state of affairs here is well outlined in e.g. \cite{Bojo12}.]  

\mbox{ } 

\noindent 7) The {\bf Spacetime Reconstruction Problem} from space and/or a discrete ontology as per Fig 2.f) is further motivated at the quantum level as per Fig 2.j).
 
\mbox{ } 

\noindent 8) {\bf Global Problems of Time}.  
These are difficulties with choosing an 'everywhere-valid' timefunction, and further global holes in other Problem of Time facets.
\noindent A couple of particular specifically-quantum global issues with which there has been some recent progress \cite{Fashion} are as follows (see \cite{AGlob} for more).  

\noindent a) What does it mean to patch together unitary evolutions?

\noindent b) Are quantum-level beables/observables defined everywhere, and, if not, how does one patch between different sets of these?  

\mbox{ } 

\noindent 9) {\bf Multiple Choice Problem}.          
This is only relevant once the quantum level is under consideration.  
I.e. as per Fig 2.l), canonical equivalence of classical formulations of a theory does not imply unitary equivalence of the quantizations of each \cite{Gotay}.  
By this, different choices of timefunction can lead to inequivalent quantum theories.

\noindent Postulate 9) (Uniqueness Specifications) for Background Independent schemes would then be that these are to be quantum-mechanically unambiguous.
[More precisely, there should only be physically meaningful ambiguities.]

\section{Conclusion}

\noindent I have characterized Background Independence in terms of three types of Relationalism (Temporal, Configurational and Spacetime), Constraint Closure, Expression in terms of 
Beables and some further properties of GR spacetime which include Refoliation Invariance.   
This account in no way fully covers Globality and Uniqueness Specifications, though I have also outlined some of these. 
This characterization is rather more selective than merely demanding the first two types of Relationalism, 
which is a positive feature as regards Background Independence being a filter of theories.
E.g. geometrodynamics and Loop Quantum Gravity pass the filter. 
I have also exposited the Problem of Time as a {\sl consequence} of the physically, philosophically and conceptually desirable option of directly implementing (metric) 
Background Independence. 
[This is to be contrasted this with pragmatic statements to avoid the Problem of Time by having one's theory carry some types of background dependence.  
Nonetheless, I emphasize that such models {\sl do} remain useful as examples that carry some aspects of Background Independence and not others.
This isolation of various aspects has aided in the present paper's quest to {\sl conceptually and mathematically characterize} Background Independence.]  

\noindent Example 1) Relational mechanics is less background independent than GR due to having no notion of refoliation and its invariance. 

\noindent Example 2) So are Einstein-Aether Theory and Ho\v{r}ava--Lifshitz Theory (and Causal Dynamical Triangulation via its giving Ho\v{r}ava--Litshitz Theory as its classical limit 
\cite{CDT}), due to their being foliation-dependent.   

\noindent Example 3) Scalar gravity (Nordstr\"{o}m, Einstein--Fokker theories) has an absolute causal structure, see e.g. p 28 of \cite{KieferBook} 
(as well as failing the light deflection test, by which they are physically overruled).

\noindent Example 4) Supersymmetry cannot be considered at the classical level within the paradigm given here.  
In fact, classical fermions cannot be considered. 
[These are moreover well-known to be questionable entities at the classical level on the further grounds of possessing inherently quantum-mechanical properties.] 
The reason that the above cannot be considered is that, whilst they can be cast in Leibnizian form, fail to have an `all change' or `STLRC' Machian resolution of Temporal Relationalism.  
I.e. classical fermionic change cannot enter the expression for the emergent classical Machian time.  
Thus the relational status of fermions (and consequently of supergravity and M-Theory) need to be studied specifically at the quantum level, 
which is beyond what the current article can cover.  

\noindent Example 5) Perturbative strings and covariant perturbative quantization of gravity are also excluded on grounds of possessing background metric structures.  

\mbox{ }  

\noindent Thus two major future directions for this work are considering how one can relationally formulate each of a) supersymmetry specifically at the quantum level and 
b) theories possessing extended objects (such as strings or branes).  
A third future direction concerns semiclassical and fully quantum understanding of whether and how Constraint Closure, Refoliation Invariance and Spacetime Reconstruction 
of classical GR carry over.
A fourth future direction is considering the classical and quantum discrete counterpart of the current paper. 
E.g. \cite{Dittrich} has taken a number of useful steps in this direction.  
[Spacetime Reconstruction becomes more formidable in many such approaches, and the status of diffeomorphism invariance is less well studied than in the continuum case.]  

\mbox{ } 

\noindent A remaining caveat is that this paper's detailed account is solely a local theory of {\sl metric-level} Background Independence.
Topological Background Independence is a more advanced and much more highly undeveloped subject.
A few tentative steps taken so far in this direction are as follows.

\noindent i) An extension of Loop Quantum Gravity based not on holonomies but on monodromies has been suggested \cite{Duston}.

\noindent ii) One could instead attempt a spectral approach to geometry such as in \cite{Connes}.  

\mbox{ } 

\noindent{\bf Acknowledgements}. I thank those close to me for support.  
I thank Flavio Mercati for some help with making Figure 2. 
Marc Lachieze-Rey, Malcolm MacCallum, Claus Kiefer, Don Page, Reza Tavakol, Juan Valiente-Kroon and Jeremy Butterfield for help with my career.
Harvey Brown and Simon Saunders for inviting me to speak on these matters at Oxford. 
Many of the above, Sean Gryb and Julian Barbour for discussions.
Grant FQXi-RFP3-1101 from the Foundational Questions Institute (FQXi) Fund, administered by 
Silicon Valley Community Foundation, Theiss Research and the CNRS, hosted with Marc Lachieze-Rey at APC.

\begin{appendices}

\section{Supporting Table of Figures}

{            \begin{figure}[ht]
\centering
\includegraphics[width=1.0\textwidth]{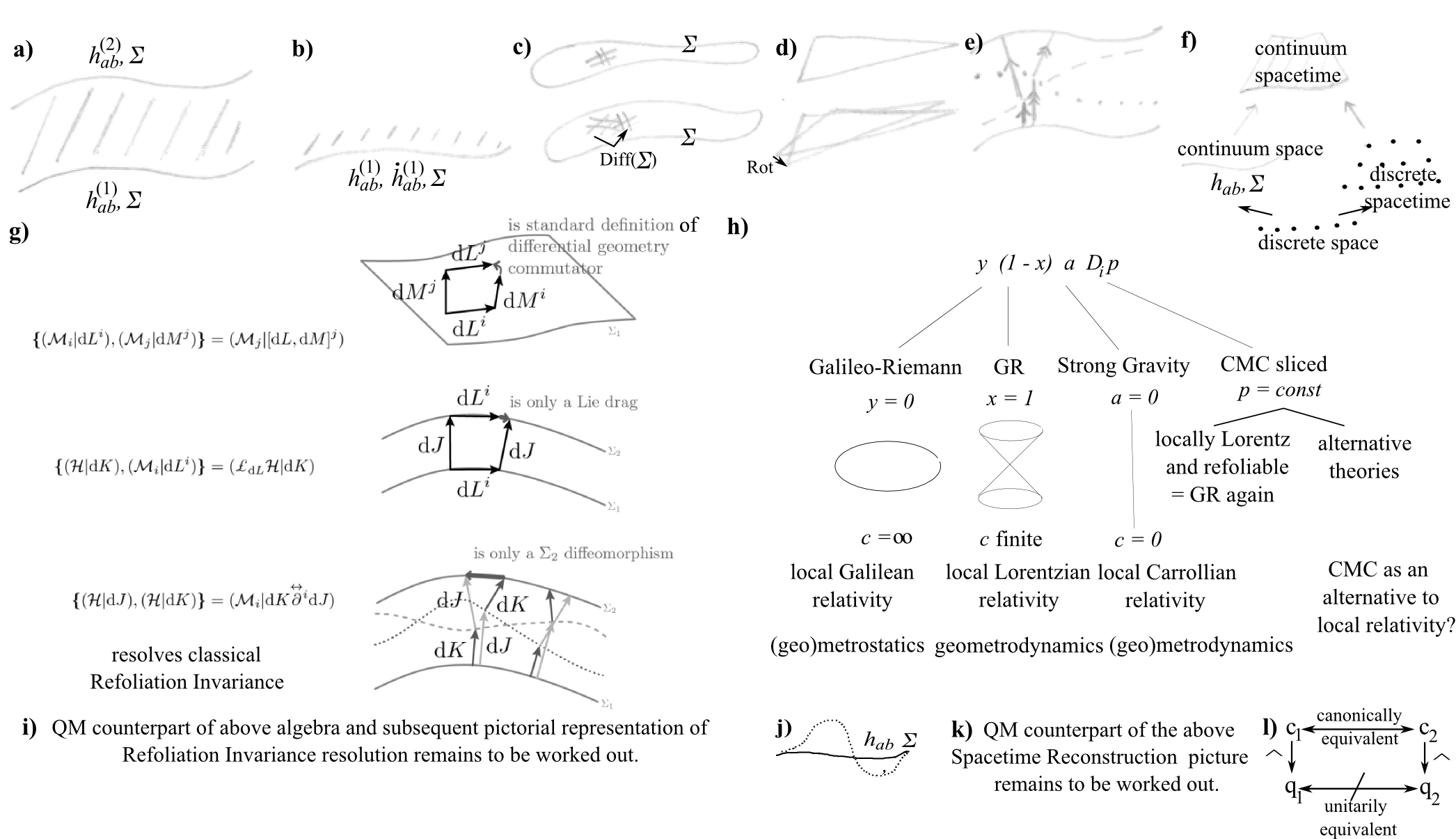}
\caption[Text der im Bilderverzeichnis auftaucht]{        \footnotesize{a) Thick sandwich bounding bread-slice data to solve for the spacetime `filling'. 
This failed to be well-posed. 
b) It was succeeded by Wheeler's idea of Thin Sandwich data to solve for a local coating of spacetime \cite{Wheeler63}.
c) The Thin Sandwich can then be re-interpreted in terms of Best Matching Riem($\Sigma$) with respect to Diff($\Sigma$), which amounts to the shuffling as depicted.  
d) The analogous triangleland Best Matching for relational triangles with respect to the rotation group, Rot.
This depiction symbolizes that Best Matching is universal \cite{BB82} rather than solely tied to geometrodynamics or the diffeomorphisms corresponding to that. 
e) Figurelet supporting the definition of classical Refoliation Invariance in GR.
f) Pictorial form of 3 types of Spacetime Reconstruction: spacetime from space, from discrete spacetime and from discrete space. 
g) Pictorial form of Dirac algebroid, including Teitelboim's classical resolution of Refoliation Invariance \cite{T73}. 
h) The outcome of classical `spacetime from space' reconstruction, as per \cite{RWR, AM13}.  
The top line of this figurelet is the obstruction term that arises from the bracket of the following generalization of the Hamiltonian constraint with itself: 
$y\{h^{ik}h^{jl} - xh^{ij}h^{kl}\}/\sqrt{h} - \sqrt{h}\{aR + b\}$. 
$c$ is the resulting theory's propagation speed. 
i) and k) reflect that we would like quantum-level counterparts of Figs g) and h) but these are not yet available.
j) depicts Wheeler's \cite{Battelle} additional subfacet of spacetime reconstruction from GR's spatial configurations at the quantum level.
I.e. that the fundamental dynamical entities -- the configurations -- then quantum-mechanically fluctuate, and, in doing so, in general cease to fit inside spacetime.
l) depicts the Multiple Choice Problem as an algebraic impasse (non-commutation of the indicated diagram of maps).  
Here c stands for classical formulation, q for quantum formulation and $\widehat{\mbox{ }}$ denotes `quantization map'.
} }
\label{Facet-Intro} \end{figure}          }

\end{appendices}


\end{document}